\documentclass[DIV16,twocolumn,10pt,headsepline,abstract=true]{scrartcl}
\usepackage[english]{babel}
\usepackage[utf8]{inputenc}
\usepackage[intlimits]{amsmath}
\usepackage{graphicx}

\usepackage{booktabs}
\usepackage{xspace}
\usepackage{url}
\usepackage{subfig}
\usepackage{color}
\usepackage[expproduct=cdot,alsoload={hep,binary}]{siunitx}
\addtokomafont{pagehead}{\small}
\addtokomafont{section}{\normalsize}
\addtokomafont{subsection}{\normalsize\normalfont}

\newcommand{\we}{{\text{w}}} 
\newcommand{\nw}{{\text{n}}} 
\newcommand{\wc}{{\text{a}}} 
\newcommand{\nc}{{\text{b}}} 

\newcommand{\co}{\ensuremath{\text{CO}_{\text{2}}}\xspace}

\newunit{\years}{\text{year}}

\title{Modeling and Simulation of Two-Phase Two-Component Flow with Disappearing Nonwetting Phase}
\author{Rebecca Neumann \and Peter Bastian \and Olaf Ippisch}
\date{}
\publishers{\large
  Interdisciplinary Center for Scientific Computing\\
  University of Heidelberg, Germany\\\vspace{2mm}
  Email: rebecca.neumann@iwr.uni-heidelberg.de
}

\begin{document}

\maketitle

\begin{abstract}\noindent {\it
Carbon Capture and Storage (CCS) is a recently discussed new
technology, aimed at allowing an ongoing use of fossil fuels while
preventing the produced \co to be released to the atmosphere. CSS can
be modeled with two components (water and \co) in two phases (liquid
and \co). To simulate the process, a multiphase flow equation with
equilibrium phase exchange is used. One of the big problems arising in
two-phase two-component flow simulations is the disappearance of the
nonwetting phase, which leads to a degeneration of the equations
satisfied by the saturation. A standard choice of primary variables,
which is the pressure of one phase and the saturation of the other
phase, cannot be applied here.

We developed a new approach using the pressure of the nonwetting phase and
the capillary pressure as primary variables. One important advantage of this
approach is the fact that we have only one set of primary variables
that can be used for the biphasic as well as the monophasic case. We
implemented this new choice of primary variables in the DUNE
simulation framework and present numerical results for some test
cases.}

\end{abstract}

\section{Introduction}
\label{intro}
In this work we address the mathematical modeling and numerical
simulation of multiphase multicomponent flow in porous media with a
special regard to \co storage in geologic formations. Some people
consider \co storage, e.g. in deep saline aquifers, as an important
factor in the effort to reduce the emission of greenhouse gases.
Reliable simulation data is crucial for all stages of CCS projects.

After the \co injection, several different trapping mechanisms lead to
an entrapment of the \co. Shortly after the injection, structural
trapping through caprocks is the most important factor. Later
solubility trapping, where \co is dissolved into water, and
residual trapping get more important. After several thousand years,
there could also occur mineral trapping caused by geochemical
reactions, but these are not considered in this work.

The mathematical model describing \co injection in geologic reservoirs
is a two-phase two-component flow in porous media, a system of
coupled, nonlinear partial differential equations. We do not only have
two different phases (liquid and \co), but also two components (water
and \co) in each phase, as the solubility of the components in the
phases has to be taken into account. For an isothermal system we have
to choose two primary variables and additional algebraic relations to
close the system.

A standard choice for the primary variables is the pressure of one
phase and the saturation of the other phase. A great challenge in this
context is the disappearance of the nonwetting phase, which has been
studied in many recent papers, as the saturations cannot be used
as primary variable here. A valid choice in the one-phase region would
be one phase pressure and the solubility of \co in the liquid phase.

Several Approaches to treat this problem exist, Class et al.\@ (2002)
switch primary variables depending on present phases, Jaffré et
al.\@ (2010) use complementarity conditions and Abadpour et al.\@ (2009)
extend the saturation to negative values. Bourgeat et al.\@ (2010) use
liquid phase pressure and water mass concentration as primary
variables.

In this study we present a new choice of primary variables that
is valid for the monophasic as well as the biphasic case and can
easily handle the disappearance of one phase. One advantage of our
approach is, that the chosen variables are continuous over material
heterogeneities, if both phases are present.

To simulate CSS, constitutive relations between physical properties
like pressure and density are necessary. We summarize our choice of
existing approaches.

Numerical simulations for different test cases presented in this work
will show that this new approach handles various applications very
well. We use our new approach to simulate \co injection into the
subsurface. A recent benchmark from the MoMas group concentrates on
test cases arising from underground radioactive waste repository
simulations. With our new set of primary variables we can also solve
these problems efficiently.

\section{Mathematical model of a isothermal two-phase two-component flow}
\label{sec:2}
In this section we will consider a porous medium and derive a system
of partial differential equations describing two-phase two-component
flow. For the sake of simplicity we use a constant temperature in this
article, but thermodynamic effects can be included into the model in a
straightforward manner. We also assume that the salinity of the water
is constant.

\subsection{Notation}
We have two phases $\alpha \in \{\we, \nw\}$, wetting and
nonwetting, and two components $\kappa \in \{\wc, \nc\}$,
water and nonwetting component.
\begin{tabbing}
  $\rho_{\text{mass},\we}$, $\rho_{\text{mass},\nw}$ \=
  liquid and nonw. phase mass densities\kill
  $p_\we$, $p_\nw$ \> wetting and nonw. phase pressures \\
  $S_\we$, $S_\nw$ \> wetting and nonw. phase saturations\\
  $\rho_{\text{mass},\we}$, $\rho_{\text{mass},\nw}$ \>
  wetting and nonw. phase mass dens.\\
  $\rho_{\text{mol},\we}$, $\rho_{\text{mol},\nw}$ \>
  wetting and nonw. phase molar dens.\\
  $\mu_\we$, $\mu_\nw$ \>
  wetting and nonw. phase viscosities\\
  $x_\we^\wc$, $x_\we^\nc$ \> molar fraction of
  comp. in wet. phase\\
  $x_\nw^\wc$, $x_\nw^\nc$ \> molar fraction of
  comp. in nonw. phase\\
  $M^\wc$, $M^\nc$ \> molar mass of wet.
  and nonw. comp. \\
\end{tabbing}

\subsection{Darcy's law}
The phase velocities $u_{\alpha}$ are given by an extended Darcy's Law:
\begin{align}\label{eq:1}
  u_\we &= -K\frac{k_{\text{r}\we}(S_\we)}{\mu_\we}
  (\nabla p_\we - \rho_{\text{mass},\we} \cdot g),\\
  \label{eq:2}
  u_\nw &= -K\frac{k_{\text{r}\nw}(S_\nw)}{\mu_\nw}
  (\nabla p_\nw - \rho_{\text{mass},\nw} \cdot g),
\end{align}
where $K$ is the absolute permeability, $k_{\text{r}\we}$ and
$k_{\text{r}\nw}$ denote the relative permeability functions and $g$
is the gravity vector.

The phase saturations and molar fractions satisfy
\begin{equation}\label{eq:3}
S_\nw  + S_\we = 1, \quad x_\we^\wc + x_\we^\nc = 1,
\quad x_\nw^\wc + x_\nw^\nc = 1.
\end{equation}
The relation between the phase pressures is given through the
capillary pressure by the Brooks-Corey or van Genuch\-ten-Mualem model
\begin{equation}\label{eq:4}
 p_{\text{c}}(S_\we) = p_\nw - p_\we.
\end{equation}

\subsection{Diffusive flux}
Following Fick's Law, the diffusive flux of a component $\kappa$ in
the phase $\alpha$ is given by
\begin{equation}\label{eq:5}
  j_{\alpha}^{\kappa} = -D_{\text{pm},\alpha}^{\kappa} \, \rho_{\text{mol},\alpha}
  \nabla x_{\alpha}^{\kappa},
\end{equation}
where $D_{\text{pm},\alpha}^{\kappa}$ is the diffusion coefficient of component
$\kappa$ in phase $\alpha$ in a porous medium.

Like \cite{class_2000} and \cite{bourgeat_2010} we assume
\begin{equation}\label{eq:6}
j_{\alpha}^\wc + j_{\alpha}^\nc = 0
\end{equation}
holds for simplicity, so we only need two diffusion coefficients
instead of four.

\subsection{Mass conservation}
Local equilibrium phase exchange of the components in the phases is assumed.
Taking into account the conservation of the amount of substance of
each component and using \eqref{eq:1}, \eqref{eq:2} and \eqref{eq:5}
we get the following partial differential equations describing an
isothermal two-phase two-component flow:
\begin{alignat}{4}\label{eq:7}
  \phi \partial_t \{ \rho_{\text{mol},\we} \, x_\we^\wc S_\we
  &+ \rho_{\text{mol},\nw} \, x_\nw^\wc S_\nw \}&& \notag\\
  &+ \nabla \cdot \{ \rho_{\text{mol},\we} \, x_\we^\wc u_\we 
  &+ &\, \rho_{\text{mol},\nw} x_\nw^\wc u_\nw \} \notag\\
  &+ \nabla \cdot \{ j_\we^\wc + j_\nw^\wc \}
  &- &\,q^\wc = 0, \notag\\
  \phi \partial_t \{ \rho_{\text{mol},\we} \, x_\we^\nc S_\we
  &+ \rho_{\text{mol},\nw} \, x_\nw^\nc S_\nw \}&& \notag\\
  &+ \nabla \cdot \{ \rho_{\text{mol},\we} \, x_\we^\nc u_\we
  &+ &\, \rho_{\text{mol},\nw} x_\nw^\nc u_\nw \} \notag\\
  &+ \nabla \cdot \{ j_\we^\nc + j_\nw^\nc \}
  &- &\,q^\nc = 0,
\end{alignat}
where $q^\wc$ and $q^\nc$ are the source/sink terms
for the components.

\section{Constitutive Relations}
\label{sec:3}
We will now look at the special case of CCS where a liquid water phase
and a liquid, gaseous or supercritical \co phase are present. The
components are water and \co. In the following the different functions
and Equations of State (EOS) to determine secondary parameters are
described. Additionally their dependence on other variables is given.

\subsection{Solubility of components}
\label{sec:3.1}
\begin{figure}[ht]
  \centering
  \includegraphics[width=0.45\textwidth]{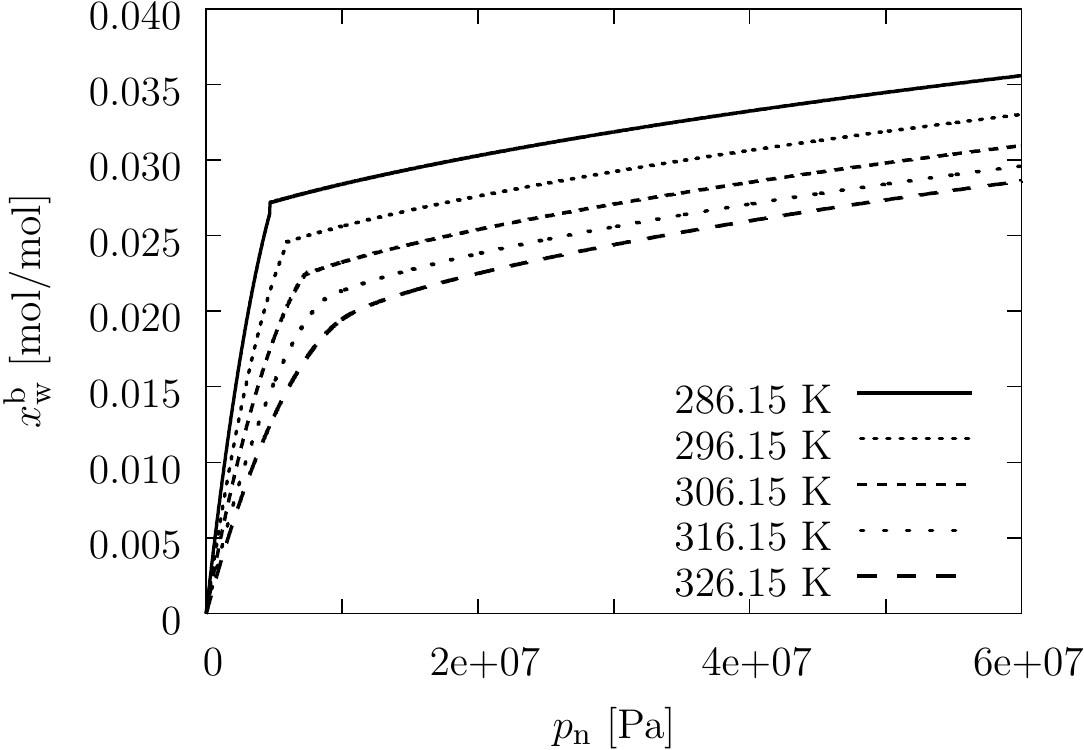}
  \caption{Solubility $x_\we^\nc$ for different temperatures
    ($s_{\text{sal}}=0$)}
  \label{fig:1}
\end{figure}

\begin{figure}[ht]
  \centering
  \includegraphics[width=0.45\textwidth]{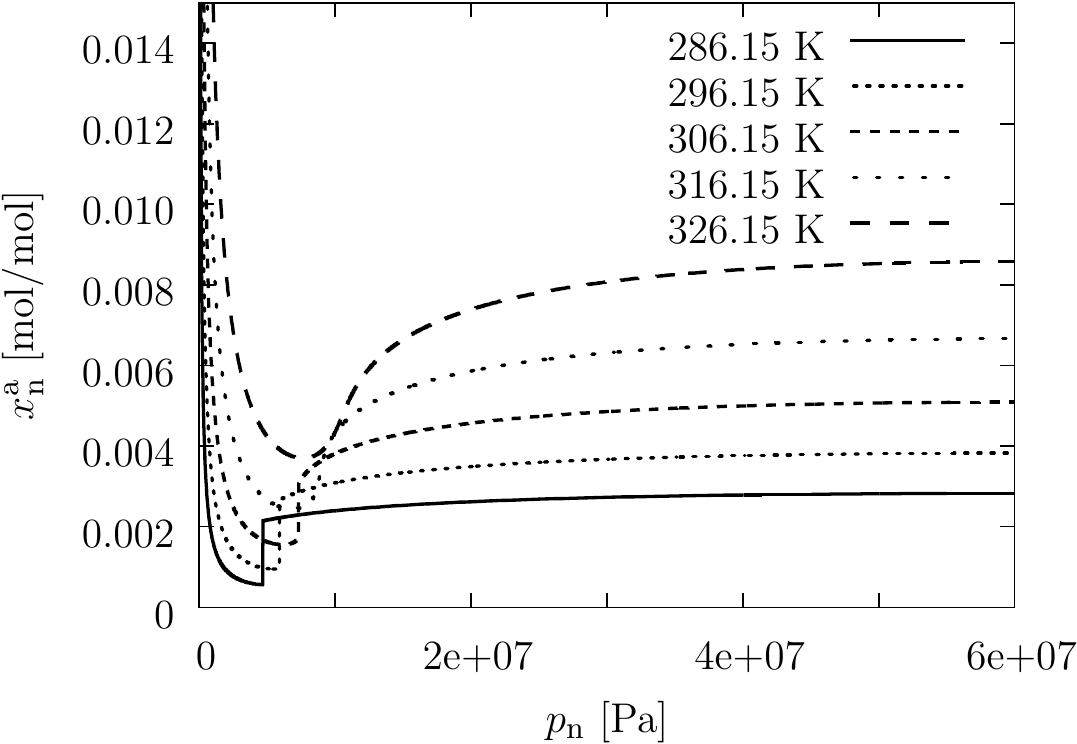}
  \caption{Solubility $x_\nw^\wc$ for different temperatures
    ($s_{\text{sal}}=0$)}
  \label{fig:2}
\end{figure}

The solubility of the components is influenced by the pressure $p_\nw$
and the temperature $T$ of the system, the salinity $s_{\text{sal}}$
of water also plays an important role:
\begin{equation}\label{eq:8}
x_\we^\nc(p_\nw,T,s_{\text{sal}}), \quad x_\nw^\wc(p_\nw,T,s_{\text{sal}}).
\end{equation}
There exist different EOS for this system. We use the EOS by Spycher
\& Pruess \cite{spycherpruess_2005}, because in contrast to other
models (for example, the EOS of Duan \& Sun \cite{duansun_2003}) also
the solubility of water in \co is described very well. Figure
\ref{fig:1} and \ref{fig:2} show the solubility curves for different temperatures.

The solubility of \co in the water phase increases fast with rising
pressure up to the saturation pressure, above that it rises with a
smaller rate. For temperatures below the critical temperature
$T_{\text{crit}}=\SI{304.15}{\kelvin}$, the state of the carbon
dioxide changes from gaseous (below saturation pressure) to liquid
which results in a not continuously differentiable sharp break at the
transition point.
\subsection{Densities}
\begin{figure}[ht]
  \centering
    \includegraphics[width=0.45\textwidth]{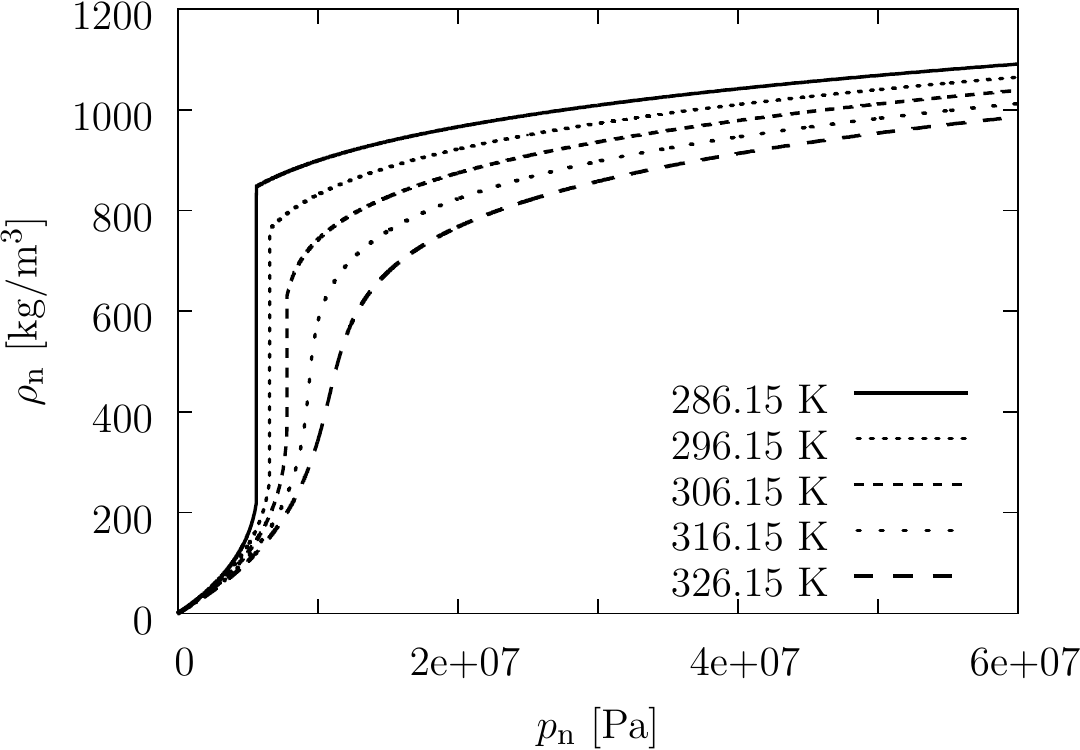}
  \caption{ \co density for different temperatures}
  \label{fig:3}
\end{figure}

For the density of the water phase the approach of Garcia
\cite{garcia_2001} is applied. The density increases slightly for a
larger fraction of \co in the water phase. The EOS of Duan
\cite{duan_1992} is used to calculate the density of the \co phase,
which strongly depends on the \co phase pressure,
\[ \rho_{\text{mass},\we}(x_\we^\nc,T), \quad \rho_{\text{mass},\nw}(p_\nw,T). \]
Figure \ref{fig:3} shows the density of \co for different
temperatures.  To convert mass density to molar density the phase
composition has to be taken into account,
\[ \rho_{\text{mol},\alpha} = \frac{\rho_{\text{mass},\alpha}}
   {x_\alpha^\nc M^\nc + x_\alpha^\wc M^\wc}. \]
\subsection{Viscosities}
\begin{figure}[ht]
  \centering
    \includegraphics[width=0.45\textwidth]{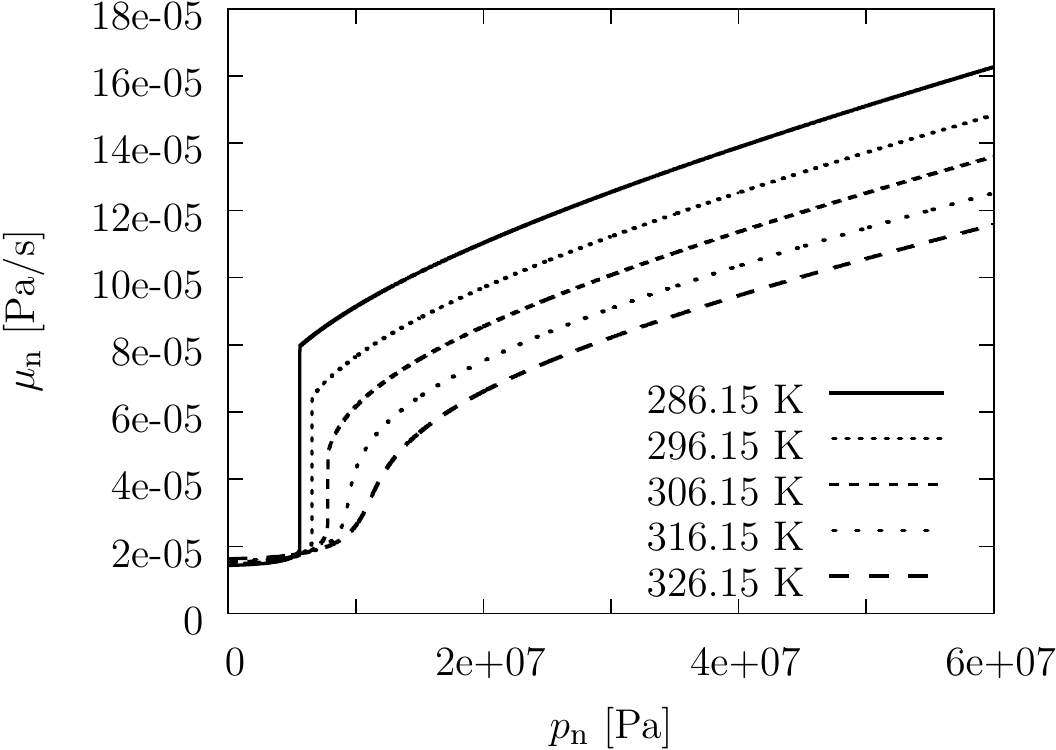}
  \caption{\co viscosity for different temperatures}
  \label{fig:4}
\end{figure}

The viscosity of the water phase is computed with a function from
Atkins \cite{atkins_1990}, for the \co phase we use the approach of
Fenghour \& Vesovic \cite{fenghour_1998}. Again the \co phase
viscosity strongly depends on the \co phase pressure,
\[ \mu_\we(T), \quad \mu_\nw(p_\nw,T). \]
Figure \ref{fig:4} shows the viscosity of \co for different temperatures.
\subsection{Diffusion}
Following \cite{jin_1996}, we use an approach suggested by Millington
\& Quirk
\[ D_{\text{pm},\alpha}^{\kappa} = 
\frac{(\phi S_{\alpha})^{10/3}}{\phi^2} D_{\alpha}^{\kappa}, \]
for the diffusion coefficient in the porous medium, where
$D_{\alpha}^{\kappa}$ describes the binary diffusion coefficient of
component $\kappa$ in phase $\alpha$.

\section{Choice of primary variables}
\label{sec:4}
A standard choice for the primary variables are one phase pressure and
the saturation. In the one phase region ($S_\nw=0$), the system \eqref{eq:7}
degenerates to
\begin{alignat*}{10}
  \phi \partial_t \left\{ \rho_{\text{mol},\we} \, x_\we^\wc \right\}
  &+ \nabla \cdot \left\{ \rho_{\text{mol},\we} \, x_\we^\wc
  u_\we + j_\we^\wc \right\} &&-\, q^\wc &&= 0,\\
  \phi \partial_t \left\{ \rho_{\text{mol},\we} \, x_\we^\nc \right\}
  &+ \nabla \cdot \left\{ \rho_{\text{mol},\we} \, x_\we^\nc \,
 u_\we + j_\we^\nc \right\} &&-\, q^\nc &&= 0.
\end{alignat*}
Using \eqref{eq:3} and \eqref{eq:6} the system can be rewritten as a
coupled groundwater-flow and transport problem
\begin{alignat}{3} \label{eq:9}
  &\phi \partial_t \left\{ \rho_{\text{mol},\we} \right\}
  &&+ \nabla \cdot \left\{ \rho_{\text{mol},\we} u_\we \right\}
  = q^\wc + &&q^\nc,\\
  &\phi \partial_t \left\{ \rho_{\text{mol},\we} \, x_\we^\nc \right\}
  &&+ \nabla \cdot \left\{ \rho_{\text{mol},\we} \, x_\we^\nc \,
 u_\we + j_\we^\nc \right\} = &&q^\nc. \notag
\end{alignat}
With the disappearance of the nonwetting phase the saturation can no longer
be used as primary variable and the standard choice of variables
cannot be applied here. One natural set of variables for the one
phase region would be $p_\we$ and $x_\we^\nc$.
\begin{table}[ht]
  \centering
  \begin{tabular}{l@{\em}r@{}}
    \toprule
    Primary Var. & Method \\
    \midrule
    $S_ \nw$, $p_\nw$  & Extending the saturation\\
    & to negative values (see \cite{abadpour_2009}).\\
    $S_\we$, $p_\we$, $X_\we^\nc$ & Using
    complementarity constraints\\
    & (see \cite{jaffre_2009}).\\
    $p_\nw$, ($S_\we$ or $X_\we^\nc$)
    & Switching primary variables depen-\\
    & ding on present phases (see \cite{forsyth_1991}, \cite{classhelmig_2002}).\\
    \bottomrule
  \end{tabular}
  \caption{Several methods to deal with a disappearing nonwetting phase}
  \label{tab:1}
\end{table}
There are several approaches to solve the problem at the phase
transition (see Table \ref{tab:1}, $X_\alpha^\kappa$ denotes the mass fraction).

A common method is primary variable switching used for example by
Forsyth \& Simpson \cite{forsyth_1991} and Helmig \& Class
\cite{classhelmig_2002}. Here different sets of primary variables are
used in the one phase and two phase region, the variables are switched
if a phase appears or disappears.

Abadpour \& Panfilov \cite{abadpour_2009} extend the saturation to
artificial negative values, so that system \eqref{eq:7} does not
degenerate in the one phase region and the saturation can still be
used as a primary variable.

Jaffré \& Sboui \cite{jaffre_2009} use the solubility as an
additional third primary variable. Additional nonlinear
complementarity constraints, which describe the transition from one
phase to two phase region are used to close the system.

We developed a new approach using the pressure of the nonwetting phase and
the capillary pressure as primary variables. In the absence of the
nonwetting phase, $p_\nw$ is defined as the corresponding pressure to the
solubility $x_\we^\nc$.

Our approach has the advantage, that we only have two primary
variables in contrast to the complementarity constraints method, where
an additional variable is needed. With our constant set of variables
we also avoid a switching of the primary variables, which is a
non-differentiable process that can lead to numerical difficulties.

This idea was first presented by Ippisch \cite{ippisch_2003}.  In the
context of nuclear waste management for the special case that Henry's
Law is used to couple solubility and pressure there exist similar
approaches. Bourgeat et al.\@ \cite{bourgeat_2010} use the water mass
concentration and the wetting phase pressure, Angelini et
al.\@ \cite{angelini_2011} use the two phase pressures as primary
variables.

In section \ref{sec:5} we will apply our approach not only to a recent
benchmark study on nuclear waste management, but also to the very
challenging field of CCS. In contrast to nuclear waste management and
the work of Bourgeat et al.\@ and Angelini et al.\@ it is not possible to
use Henry's Law for the solubility, because the approximation is not
valid for \co. We need a nonlinear function (see subsection
\ref{sec:3.1}) to describe the dependency between the nonwetting
pressure and mole fraction.  Moreover we have to handle the very high
injection rate of the \co.

\subsection{$p_\nw/p_c$ formulation: Interpretation as algebraic transformation}
The entry pressure $p_{\text{entry}}$ is the critical capillary
pressure that must be applied so that the nonwetting phase appears.
We have to distinguish between
\begin{itemize}
\item[1.] $p_{\text{c}} \le p_{\text{entry}}$ where
$S_\nw=0$ and only the wetting phase exists
\item[2.] $p_{\text{c}} > p_{\text{entry}}$ where $S_\nw>0$ and both wetting and
nonwetting phase exist.
\end{itemize}

\subsection*{Case 1: $p_{\text{c}} \le p_{\text{entry}}$}
As mentioned in the beginning of the section the natural set of
variables for the one phase system \eqref{eq:9} would be $p_\we$ and
$x_\we^\nc$. Consider the following transformation of variables
\begin{align}\label{eq:10}
  p_\we &= p_\nw - p_{\text{c}}\\
  x_\we^\nc &= \psi(p_\nw) \notag
\end{align}
where $\psi$ is a continuous and invertible function. The solubility
relation for $x_\we^\nc$ \eqref{eq:8} satisfies these demands (see Figure
\ref{fig:1} and Spycher \& Pruess \cite{spycherpruess_2005}). The
mapping between $p_\nw$ and $x_\we^\nc$ is hence unique and $p_\nw$
and $p_{\text{c}}$ is a valid set of primary variables.

The relation between the capillary pressure and the saturation
$p_{\text{c}}(S_\we)$ (see Equation \eqref{eq:3}) is a strictly
decreasing function for $S_\we \in [0,1]$ and can therefore be
inverted
\[ S_\we = \eta(p_{\text{c}}). \]
The dependent variables are then obtained through
\begin{alignat*}{5}
  S_\we &= \eta(p_{\text{c}}) \quad S_\nw &&= 1 - \eta(p_{\text{c}})\\
  x_\we^\nc &= \psi(p_\nw) \quad x_\we^\wc &&= 1 - \psi(p_\nw)\\
  x_\nw^\wc &= \gamma(p_\nw) \quad x_\nw^\nc &&= 1 - \gamma(p_\nw)
\end{alignat*}
where $\gamma$ is the solubility curve given in \eqref{eq:8}.
All other variables are computed as given in Section \ref{sec:3}.

This choice is not unique, another possible set would be $p_\we/p_c$
or $p_\we/p_\nw$. Using $p_\nw$ as a primary variable has the
advantage, that the highly nonlinear density and viscosity functions
are directly dependent on a primary variable. We prefer $p_c$ over
$p_l$ as additional primary variable, because then the saturation only
depends on the primary variable $p_c$ through the nonlinear capillary
pressure-saturation relationship.

Instead of the nonwetting phase pressure $p_\nw$ the molar fraction
$x_\we^\nc=\psi(p_\nw)$ could also be used as primary variable, which is
very similar to the water mass concentration used by Bourgeat et
al. \cite{bourgeat_2010}.

\subsection*{Case 2: $p_{\text{c}} > p_{\text{entry}}$}
The common choice of primary variables in the two-phase region is one
pressure and the saturation.  With the $p_\nw/p_{\text{c}}$
formulation we obtain the saturations through the retention curve $S_\we =
\eta(p_{\text{c}})$, the other variables are computed accordingly.
\vspace{5mm}\\ $p_\we$ and $x_\we^\nc$ are continuous at the interface
between the one-phase and the two-phase region. Through the
transformation \eqref{eq:10}, $p_\nw$ and $p_{\text{c}}$ are
continuous at the interface too.  With $p_\nw/p_{\text{c}}$ we found a
set of primary variables that can be consistently used in the presence
or absence of the nonwetting phase. One advantage of the
$p_\nw/p_{\text{c}}$ formulation is, that the pressures, in contrast
to the saturations, are continuous across material heterogeneities if
both phases exist.

\section{Numerical simulation}
\label{sec:5}
In the following section we present the numerical results for special
test cases.  All simulations were performed in the {DUNE} simulation
framework \cite{dune}, \cite{dune_2008}.

A cell-centered finite volume method with two-point flux approximation
on a structured grid was used for the domain discretization.  The grid
$E_h = {e_1,\ldots, e_n}$ consists of elements $e_i$ and the boundary
of each element is $\partial e_i = \bigcup_{j \in \Sigma(i)}
\gamma_{ij}$ where $\gamma_{ij}$ denotes the boundary between elements
$e_i$ and $e_j$.  The cell-centered finite volume method for Equation
\eqref{eq:7} for each component $\kappa$ then reads
\begin{alignat*}{2}\label{eq:11}
  &\sum \limits_{e_i \in E_h} \bigg\{ \,\, \int \limits_{e_i}
  \phi \partial_t \left\{ \rho_{\text{mol},\we} \, x_\we^{\kappa} S_\we
  + \rho_{\text{mol},\nw} \, x_\nw^{\kappa} S_\nw \right\} \text{d}e && \\[-1mm]
  &\quad + \frac{1}{\|e_i\|} \sum \limits_{j \in \Sigma(i)}
  \,\, \int \limits_{\gamma_{ij}} \Big( \nabla \cdot \left\{
  \rho_{\text{mol},\we} \, x_\we^{\kappa} u_\we
  + \rho_{\text{mol},\nw} \, x_\nw^{\kappa} u_\nw \right\} &&\\[-3mm]
  & \phantom{\quad \frac{1}{\|e_i\|} \sum \limits_{j \in \Sigma(i)}
    \,\, \int \limits_{\gamma_{ij}} \Big(}
  + \nabla \cdot \left\{ j_\we^{\kappa} + j_\nw^{\kappa} \right\}
  - q^{\kappa} \Big) \cdot n_{ij} \,\, \text{d}\gamma \bigg\} = 0&&,
\end{alignat*}
where $n_{ij}$ denotes the unit outer normal to $\gamma_{ij}$.

A special upwinding scheme is used to calculate the phase fluxes 
at the interface between two elements to handle material discontinuities
resulting in different capillary-pressure
saturation curves and relative permeability functions in both elements.

The direction of the flux of phase $\alpha$ at the interface between
two elements $i$ and $j$ can be obtained from the sum of the pressure
gradient and the force of gravitation $w_{\alpha,ij} = -(\nabla
p_{\alpha} - \rho_{\text{mass},\alpha,ij} \cdot g) \cdot n_{ij}$, where
$\rho_{\text{mass},\alpha,ij}$ is computed as the arithmetic average of
cells $e_i$ and $e_j$. Depending on the sign of $w_{\alpha,ij}$ the
upwind element is determined.
\[
\text{\sffamily upwind}_{\alpha} = \begin{cases}
  i & w_{\alpha,ij}  \ge 0\\
  j & \text{else}
\end{cases}. 
\]
The capillary pressure of the upwind element is used to calculate
the relative permeability in each element. The obtained relative
permeabilities are multiplied by the absolute permeabilities and the
viscosities in each element. A harmonic average of the values is
used to calculate the flux at the interface:
\begin{align*}
K_{\alpha,i} &= K_i \frac{k_{r\alpha,i}(p_{c,\text{\sffamily upwind}_{\alpha}})}{\mu_{\alpha,i}}\\
K_{\alpha,j} &= K_j \frac{k_{r\alpha,j}(p_{c,\text{\sffamily upwind}_{\alpha}})}{\mu_{\alpha,j}}\\
u_{\alpha,ij} &= \frac{K_{\alpha,i} K_{\alpha,j}}{K_{\alpha,i}+K_{\alpha,j}} w_{\alpha,ij}
\end{align*}
For homogeneous porous media this upwinding scheme corresponds to an upwinding
of saturation.

For the calculation of the convective component transport a full
upwinding of the molar fractions and the molar densities based on the
upwind direction is used with
\[ x^{\kappa}_{\alpha,ij} = x^{\kappa}_{\alpha,{\text{\sffamily upwind}}_{\alpha}}, \quad
\rho_{\text{mol},\alpha,ij} = \rho_{\text{mol},\alpha,{\text{\sffamily upwind}}_{\alpha}}. \]

As time discretization scheme the implicit Euler Method was
used. Newton's Method was applied to linearize the system. The
Jacobian matrix is derived through numerical differentiation. The
resulting linear equation system is solved with a BiCGStab solver
preconditioned by an algebraic {mul-tigrid} method (see \cite{blatt_2010}).

We chose three different test cases, the first one is from a recent
benchmark study concentrating on appearance and disappearance of
phases in the context of nuclear waste management. As there are no
analytical solutions for two-phase two-component flow systems, we use
the results of other groups as possibility to validate our results. We
also conduct a grid convergence study to verify the experimental order
of convergence of our implementation.

With the second test case we apply our formulation to a \co sequestration
scenario in 2D and perform a strong scalability test.  The third
test case extends the second test case to 3D and shows that our
approach can handle the large number of unknowns.

The simulations were performed in parallel with up to 16 processes.

\section{Test case 1: Gas injection in a fully water saturated domain (quasi-1D)}
\label{sec:6}
The first test case is an example from the MoMas benchmark on
multiphase flow in porous media \cite{momas_benchmark_2011},
\cite{bourgeat_2012}. We converted the descriptions to match the
variables used in this paper.

\begin{figure}[ht]
  \includegraphics[width=.48\textwidth]{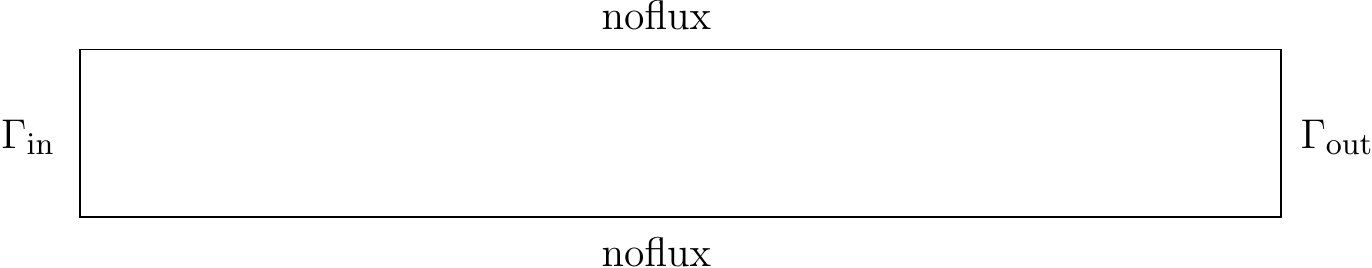}
  \caption{Test case 1: Domain setup}
  \label{fig:5}
\end{figure}

In this case the considered nonwetting component is hydrogen and the
wetting component is water. The solubility of water in the nonwetting
phase is neglected: $x_\nw^\wc = 0$. Hydrogen is injected into the
left part of a rectangular domain ($\SI{200}{\metre} \times
\SI{20}{\metre}$) with a flux of $q_\nw^{\text{in}}$ for
$\SI{5e5}{\years}$s.
\begin{table}[ht]
\centering
  \begin{tabular}{@{}l%
      @{\ }S[tabnumalign=right,tabformat=1.2e+1]%
      s[tabunitalign=left]%
      l%
      @{\ }S[tabnumalign=right,tabformat=1.2e+2]%
      s[tabunitalign=left]@{}%
    }
    \toprule
    & {Value}
    & 
    &
    & {Value}
    & \\
    \cmidrule(r){1-3}
    \cmidrule{4-6}
    $\phi$             & 0.15    &
    & $q^{\kappa}$        & 0       & \\
    $S_{\we,\text{res}}$  & 0.4     &
    & $q^{\text{in}}_{\nw}$ & 1.77e-13 & \kilogram \per \metre \squared \per \second \\
    $S_{\nw,\text{res}}$  & 0       &
    & $D_\we^\nc$         & 3e-9    & \metre \squared \per \second \\
    $n$                & 1.49    &
    & $D_\nw^\wc$         & 0       & \metre \squared \per \second \\
    $\alpha$           & 5e-7    & \per\pascal
    & $\mu_\we$          & 1e-3     & \pascal \per \second \\
    $M^\wc$             & 1e-2     & \kilogram \per \mole
    & $\mu_\nw$          & 9e-6     & \pascal \per \second \\
    $M^\nc$             & 2e-3     & \kilogram \per \mole
    & $H$                & 7.65e-6 & \mole \per \metre \cubed \per \pascal\\
    $p_\we^{\text{i}}$   & 1e6      & \pascal
    & $K$                & 5e-20    & \meter\squared  \\  
    $p_\nw^{\text{i}}$   & 0      & \pascal
    & & &\\
    \bottomrule
  \end{tabular}
  \caption{Parameters for test case 1}
  \label{tab:2}
\end{table}
The domain is initially fully saturated by the water phase,
consisting only of pure water with initial conditions $p_{\alpha} =
p_{\alpha}^{\text{i}}$ (see Table \ref{tab:2}). The boundary conditions
are Neumann 0 boundaries at the top and bottom (see Figure
\ref{fig:5}).
The Dirichlet boundary conditions for the outflow boundary are the
same as the initial conditions: 
$\left. p_{\alpha}\right|_{\mathrm{\Gamma}_{\text{out}}} = p_{\alpha}^{\text{i}}$. 
Gravitation is neglected, which leads to a quasi-1D problem.

The relationship between $p_\nw$ and $X_\we^\nc$ (where
$X_\alpha^\kappa$ is the mass fraction in contrast to the molar
fraction $x_\alpha^\kappa$) is given through Henry's Law:
\[ X_\we^\nc = \frac{H(T)}{\rho_{\text{mol},\we}} p_\nw^\nc, \]
where the partial pressure of hydrogen in the nonwetting phase is
$p_\nw^\nc = p_\nw$ for this case because there is no water in the
nonwetting phase for this example. The mass fraction $X_\we^\nc$ is
then converted to the molar fraction:
\[ x_\we^\nc = \frac{X_\we^\nc M^\wc}{X_\we^\nc M^\wc + (1-X_\we^\nc) M^\nc}. \]
The nonwetting phase density is determined by the ideal gas law,
wetting phase density is obtained through Henry's Law
\[ \rho_{\text{mass},\nw} = p_\nw M^\nc (RT)^{-1}, \quad \rho_{\text{mass},\we} =
10^3 + H(T) M^\nw p_\nc. \]
The diffusion coefficient is given as
\[ D_{\text{pm},\alpha}^{\kappa} = \phi S_{\alpha} \left( \frac{X_\alpha^\nw}{M^{\nw}}
+ \frac{X_\alpha^\wc}{M^{\wc}} \right) D_{\alpha}^{\kappa}. \]
\begin{figure}[ht]
  \includegraphics[width=.48\textwidth]{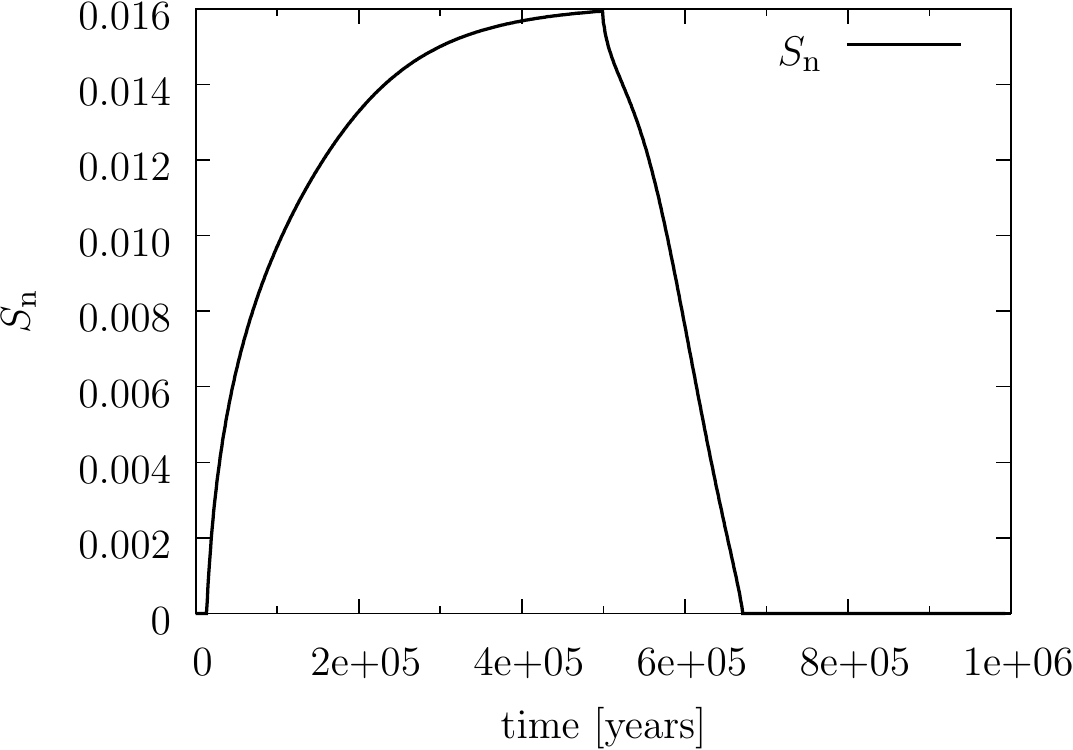}
  \caption{Test case 1: Nonwetting phase saturation at $\mathrm{\Gamma}_{\text{in}}$}
  \label{fig:6}
\end{figure}
\begin{figure}[ht]
  \includegraphics[width=.48\textwidth]{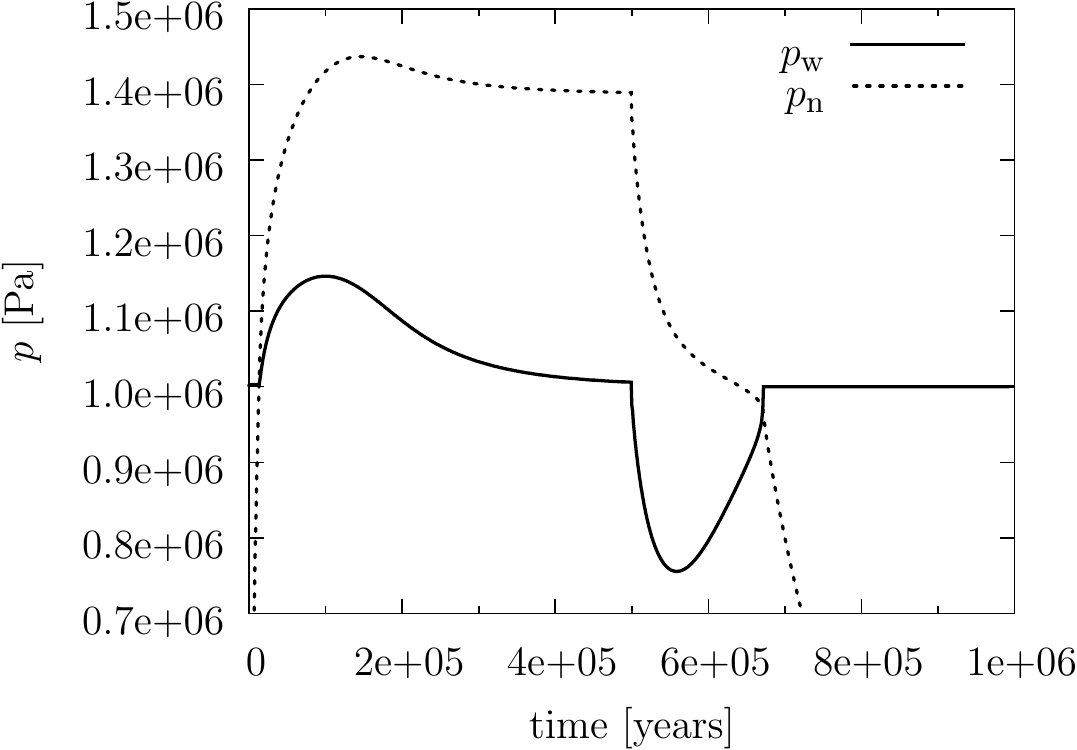}
  \caption{Test case 1: Phase pressures at $\mathrm{\Gamma}_{\text{in}}$}
  \label{fig:7}
\end{figure}
A van Genuchten-Mualem model with the parameters $n$, $\alpha$ and
$S_{\alpha,\text{res}}$ as given in Table \ref{tab:2} is used for the
soil water characteristic and relative permeabilities. All other
parameters used in the simulation are also noted in Table
\ref{tab:2}.

A structured grid with $400 \times 20$ cells was used for the computations.
Figure \ref{fig:6} and \ref{fig:7} show the nonwetting phase
saturation and phase pressures at $\mathrm{\Gamma}_{\text{in}}$ over
time. $S_\nw$ is zero at the beginning, all injected hydrogen
dissolves into the wetting phase and no nonwetting phase is present. At
$t \approx 13000$ years a nonwetting phase starts to appear at the
injection point $\mathrm{\Gamma}_{\text{in}}$.

For the computations we used a constant time step $dt = 1000$ years.
We also verified the robustness of our model by using larger time steps
($dt = 5000$ years).

Six different groups including our group participated in this
benchmark example, the results of all groups are presented in
\cite{bourgeat_2012}. The results of our simulation corresponds well
to the results of the other groups.
\begin{table}
  \centering
  \begin{tabular}{rrrr}
    \toprule
    level & \#elements & EOC ($p_c$) & EOC ($p_\nw$)\\
    \midrule
    1 & 24   & 2.01 & 2.02\\
    2 & 48   & 1.97 & 1.98\\
    3 & 96   & 1.98 & 1.98\\
    4 & 192  & 1.99 & 1.99\\
    5 & 384  & 2.00 & 1.99\\
    6 & 768  & 2.00 & 2.00\\
    7 & 1536 & 2.01 & 2.01\\
    8 & 3072 & 2.03 & 2.02\\
    \bottomrule
  \end{tabular}
  \caption{Grid convergence study for test case 1}
  \label{tab:3}
\end{table}

In addition we performed a grid convergence study. For the coarsest
level (level 1) we use $12 \times 1$ cells. For each level we double
the amount of grid cells in $x$-direction, so we have $12 \times 2^i$
cells for level $i$. The solution on level $12$ was used as a
reference solution. The resulting experimental order of convergence
(EOC) can then be computed through
\[ \text{EOC}_{i+1} = \frac{1}{\log{(2)}} | \log{ \left(\frac{e_i}{e_{i+1}}\right) } |  \]

where $e_i$ is the $L2$-error between the solution on level $i$ and
the reference solution. At $t=\SI{2e5}{}$ years we get second order
grid convergence for nonwetting phase pressure and capillary pressure
(see Table \ref{tab:3}).

The convergence study shows that our numerical solution converges with
an optimal EOC of two, which is the maximum order that can be achieved
with a cell-centered finite volume discretization.

\begin{figure*}[t]
  \centering
  \subfloat[][7 days, \quad $\max(S_\nw) = 0.82, \, \max(x_\we^\nc) = 0.022$]{
    \includegraphics[width=0.75\textwidth]{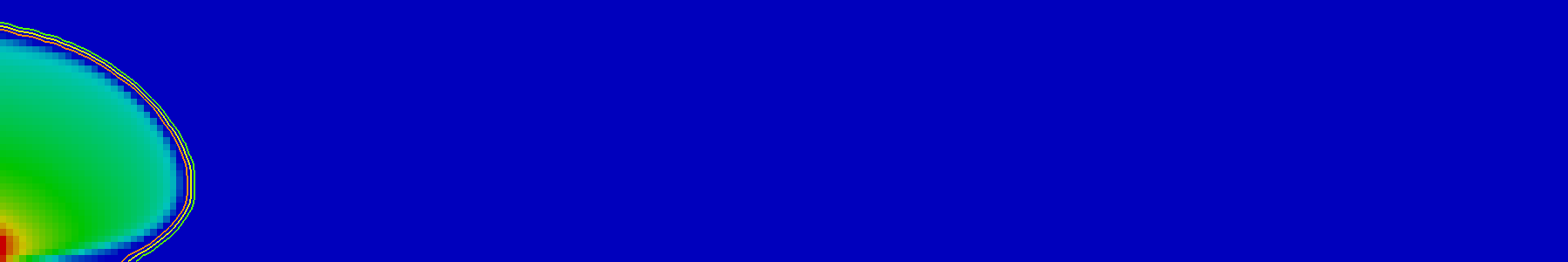}}
  \hfill
  \subfloat[][20 days, \quad $\max(S_\nw) = 1, \, \max(x_\we^\nc) = 0.020$]{
    \includegraphics[width=0.75\textwidth]{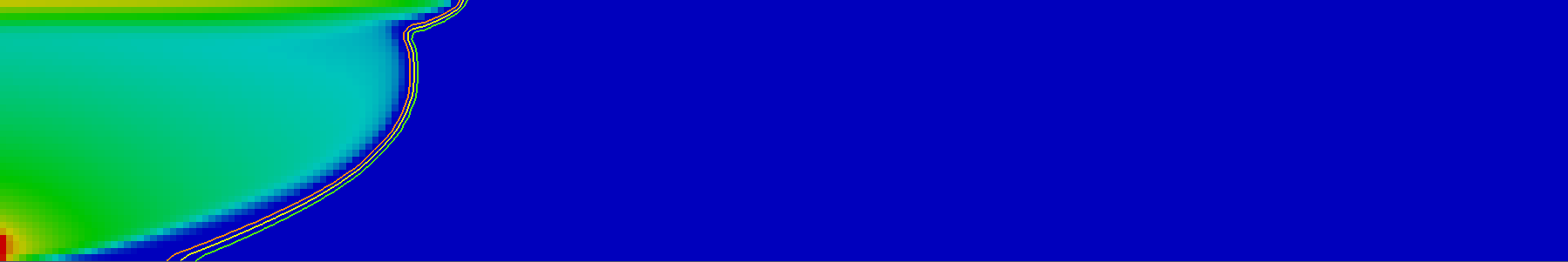}}
  \hfill
  \subfloat[][65 days, \quad $\max(S_\nw) = 1, \, \max(x_\we^\nc) = 0.020$]{
    \includegraphics[width=0.75\textwidth]{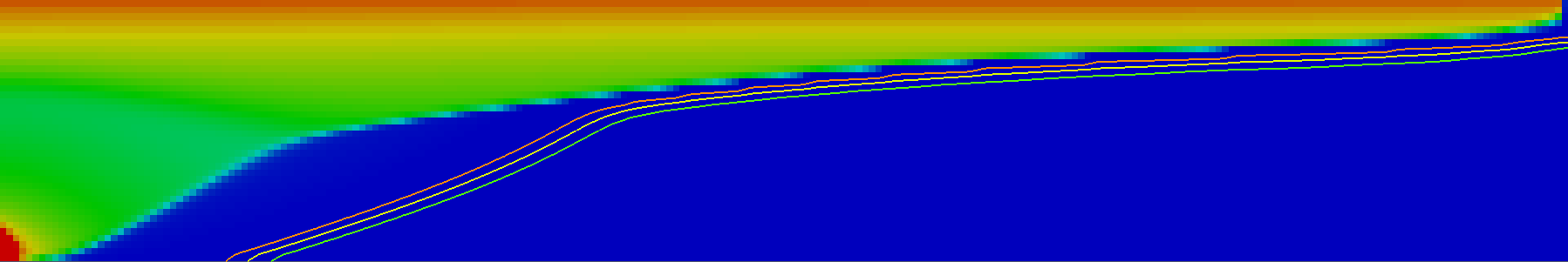}}
  \caption{Test case 2: \co phase saturation and molar fraction of
    dissolved \co in water (contour lines\\ for $x_\we^\nc=0.005$,
    $0.011$, $0.016$). Color scale ranges from $S_\nw = 0$ (blue) to
    $S_\nw = \max(S_\nw)$ (red).}
  \label{fig:8}
\end{figure*}

\section{Test case 2: \co injection in a fully water saturated domain (2D)}
\label{sec:7}
\begin{figure}[ht]
  \includegraphics[width=.48\textwidth]{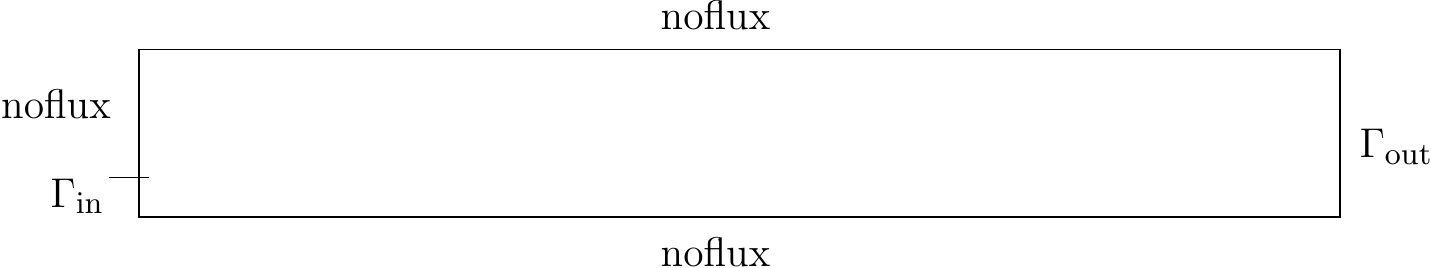}
  \caption{Test case 2: Domain setup}
  \label{fig:9}
\end{figure}

In the second test case \co is injected into the lower left part of a
rectangular geometry ($\SI{600}{\metre} \times \SI{100}{\metre}$)
with a flux of $q_\nc^{\text{in}}$. The domain is located
$\SI{800}{\metre}$ under the surface. As in test case 1, the top and
bottom of the domain have noflux boundary conditions (see Figure
\ref{fig:9}).
 For the Dirichlet boundary on the right side we choose
hydrostatic pressure for the water phase and zero pressure for the
\co phase (which leads to $x_\we^\nw = 0$)
\[\left. p_\we \right|_{\mathrm{\Gamma}_{\text{out}}} = 10^5 + (900-z)
\rho_{\text{mass},\we} \cdot g \, \, \pascal,
\quad \left. p_\nc \right|_{\mathrm{\Gamma}_{\text{out}}} = \SI{0}{\pascal}, \]
where $z$ is the $z$-coordinate in the domain and $g$ the gravity in
$z$-direction. Again the same values are taken as initial values.
\begin{table}[ht]
  \centering
  \begin{tabular}{l%
      @{\em}S[tabnumalign=right,tabformat=3.2e+2]%
      s[tabunitalign=left]%
      l%
      @{\ }S[tabnumalign=right,tabformat=1.1e+1]%
      s[tabunitalign=left]@{}%
    }
    \toprule
    & {Value}
    &
    &
    & {Value}
    & \\
    \cmidrule(r){1-3}
    \cmidrule{4-6}
    $\phi$              & 0.2    &
    & $q^{\kappa}$        & 0      & \\
    $S_{\alpha,\text{res}}$ & 0      &
    & $q^{\text{in}}_{\nw}$  & 4e-2  & \kilogram \per \metre \squared \per \second \\
    $\lambda$                 & 2      &
    & $D_\we^\nc$          & 2e-9  & \metre \squared \per \second \\
    $p_{\text{entry}}$            & 1e3 & \pascal
    & $D_\nw^\wc$          & 2e-9  & \metre \squared \per \second \\
    $K$                 & 1e-12  & \meter\squared
    & $M^\wc$             & 1.8e-2 & \kilogram \per \mole \\
    $T$                 & 313.15 & \kelvin
    & $M^\nc$             & 4.4e-2 & \kilogram \per \mole \\
    \bottomrule
  \end{tabular}
  \caption{Parameters for test case 2}
  \label{tab:4}
\end{table}

Densities, viscosities and solubilities are chosen as suggested in
Section \ref{sec:3}, all other parameters are given in Table
\ref{tab:4}. In this example we used the Brooks-Corey model for the
soil water characteristic and relative permeabilities.  For the
computations a structured grid with $240 \times 40$ cells was used.

The results of test case 2 are shown in Figure \ref{fig:8}. Each
picture shows the \co phase saturation at a specific time point, the
contour lines depict the molar fraction of \co in the water phase.
The \co migrates upwards until it reaches the top of the domain with
the noflux conditions and is then driven to the right by advective forces.
Around the \co front the water phase contains dissolved \co.

An analytical solution for this test case does not exist, but the
simulation results are plausible and the \co front behaves as
expected.

During the initial phase of CCS \co is injected and the \co front is
propagating. Thus for the sake of accuracy we want to choose a time
step size so that the \co front travels one grid cell layer per time
step.

For the computations we used a maximum time step of $dt = \SI{5000}
{\second}$. The time step size is halved if the Newton solver did not
converge, it is doubled until the maximum time step is reached in case
of convergence. With this time step control we achieve an average time
step size of $dt = \SI{3575}{\second}$ and the \co front moves about
one grid cell layer per time step, which fulfills the above condition.
\begin{table}
  \centering
  \begin{tabular}{rrr}
    \toprule
    \#processes & total time [$\second$] & efficiency\\
    \midrule
    1 & 13975 & 1\\
    2 &  7763 & 0.90\\
    4 &  4151 & 0.84\\
    8 &  2658 & 0.65\\
    \bottomrule
  \end{tabular}
  \caption{Strong scalability test for test case 2}
  \label{tab:5}
\end{table}

All simulations were done in parallel. To analyze the parallel
performance of the simulations, we conduct a strong scalability test,
where the global problem size stays fixed and the number of processes
is increased.

The efficiency is defined as
\[ E = \frac{T_1}{pT_p}, \]
where $T_1$ is the time for the sequential method, $p$ the number of
processes and $T_p$ the time for the parallel method with $p$
processes. Table \ref{tab:5} shows the results for test case 2 for a
simulation time of 65 days. The total time needed for solving the problem
scales very well with the number of processes.
\begin{table}[ht]
  \centering
  \begin{tabular}{rrrrr}
    \toprule
    \#processes & TS & av.\@ $dt$ & min.\@ $dt$ & av.\@ NI\\
    \midrule
    1 & 2249 & 3579.7 & 156.25 & 3.9\\
    2 & 2276 & 3527.6 & 156.25 & 3.9\\
    4 & 2205 & 3593.1 & 312.5 & 3.9\\
    8 & 2282 & 3514.4 & 312.5 & 3.9\\
    \bottomrule
  \end{tabular}
  \caption{Average time step size and number of Newton iterations of the strong scalability test for test case 2}
  \label{tab:6}
\end{table}

For a possible comparison with other implementations we list some
important performance indicators in Table \ref{tab:6}. TS is the
amount of time steps that were performed (successful and
unsuccessful), for average and minimum time step sizes only the
successful time steps were regarded. NI is the average number of
Newton iterations per time step (successful and unsuccessful), where
10 NI are the maximum number of iterations that were allowed. Table
\ref{tab:6} shows, that the average time step size and number of Newton
iterations stay almost constant for different number of processes. 
\begin{figure*}[ht!]
  \centering
  \subfloat[][4 days, \quad $\max(S_\nw) = 0.65, \, \max(x_\we^\nc) = 0.021$]{
    \includegraphics[width=0.34\textwidth]{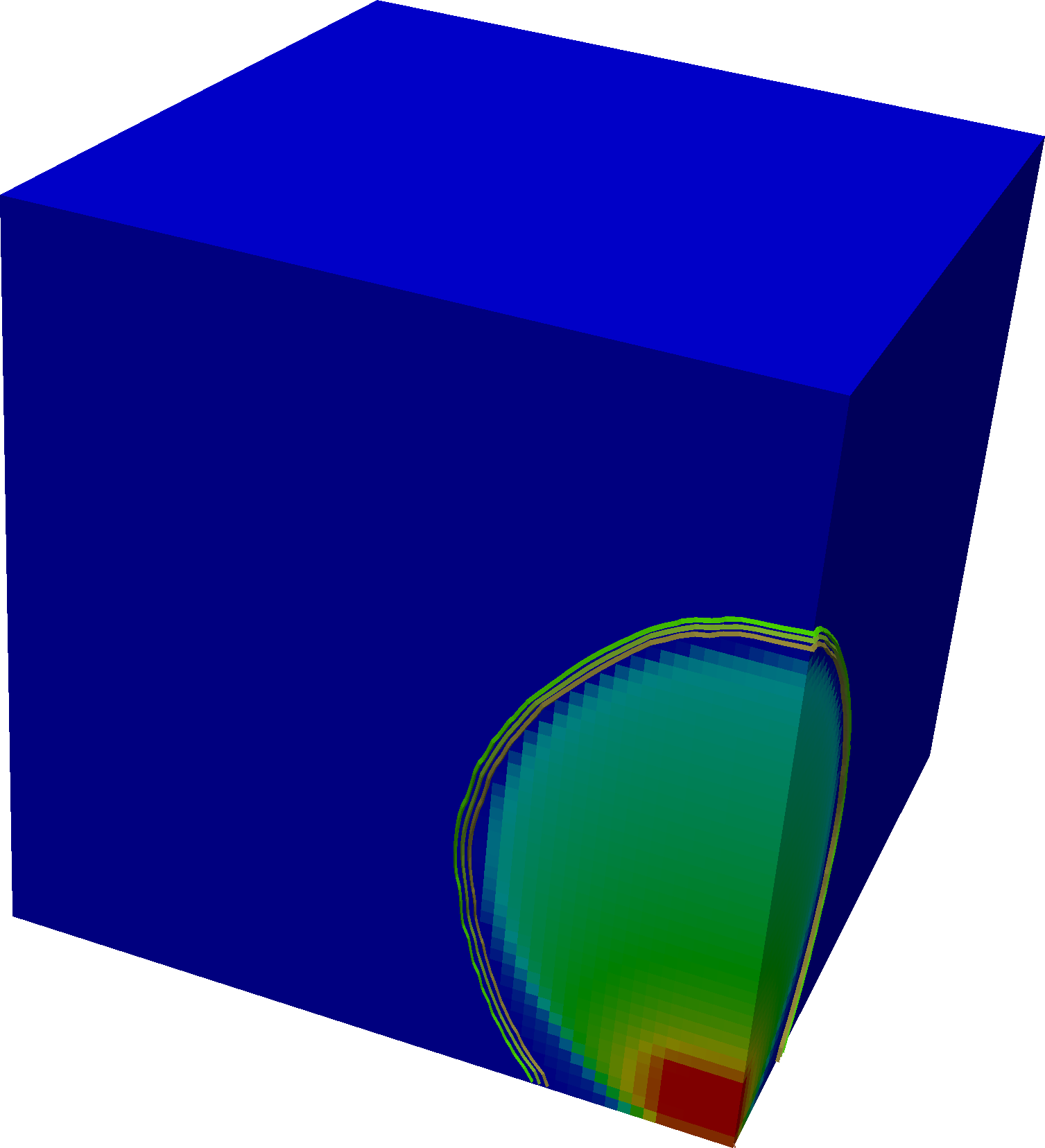}}
  \hspace{2cm}
  \subfloat[][9 days, \quad $\max(S_\nw) = 0.77, \, \max(x_\we^\nc) = 0.021$]{
    \includegraphics[width=0.34\textwidth]{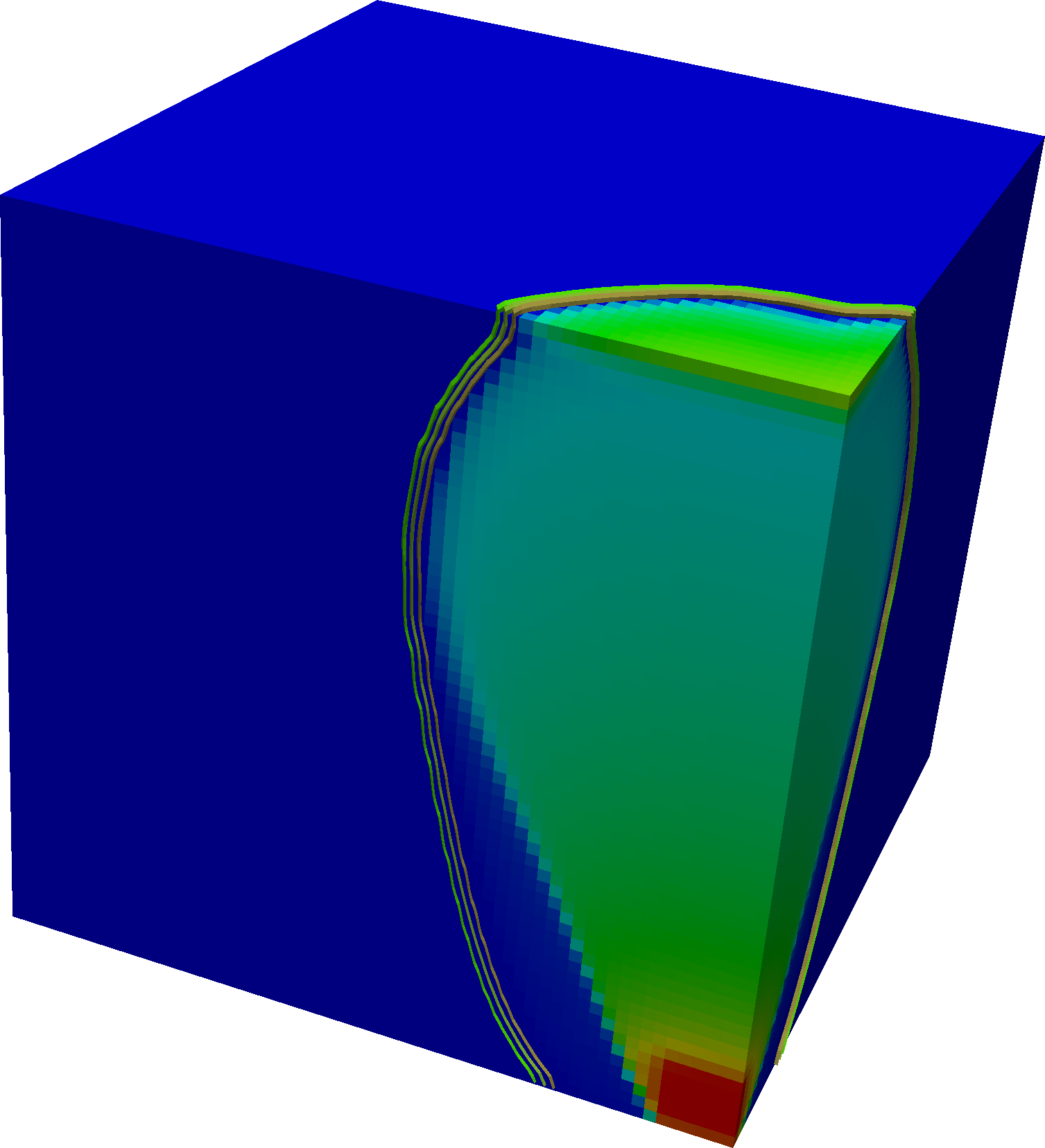}}
  \hfill
  \subfloat[][14 days, \quad $\max(S_\nw) = 0.82, \, \max(x_\we^\nc) = 0.021$]{
    \includegraphics[width=0.34\textwidth]{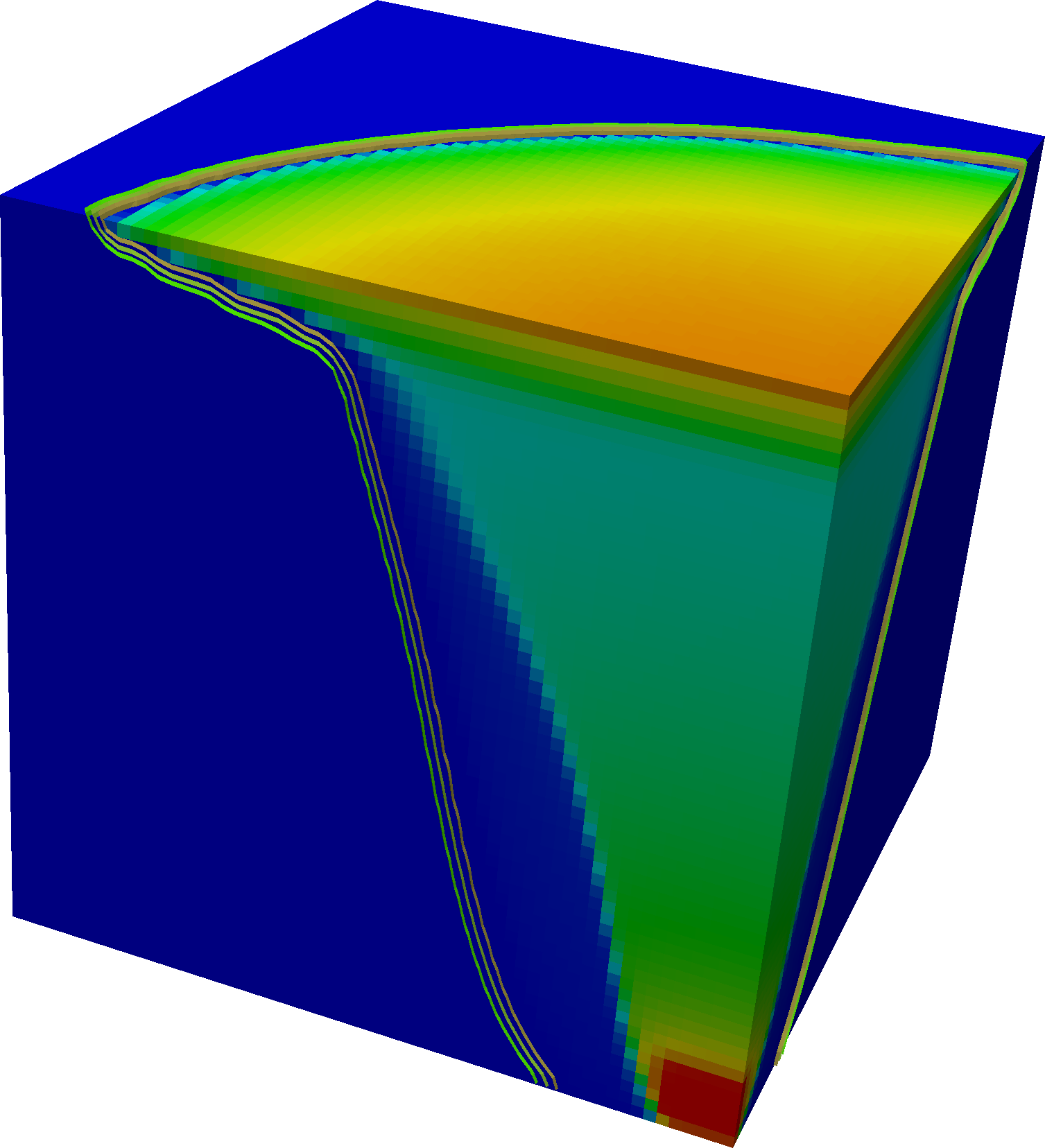}}
  \hspace{2cm}
  \subfloat[][18 days, \quad $\max(S_\nw) = 0.82, \, \max(x_\we^\nc) = 0.021$]{
    \includegraphics[width=0.34\textwidth]{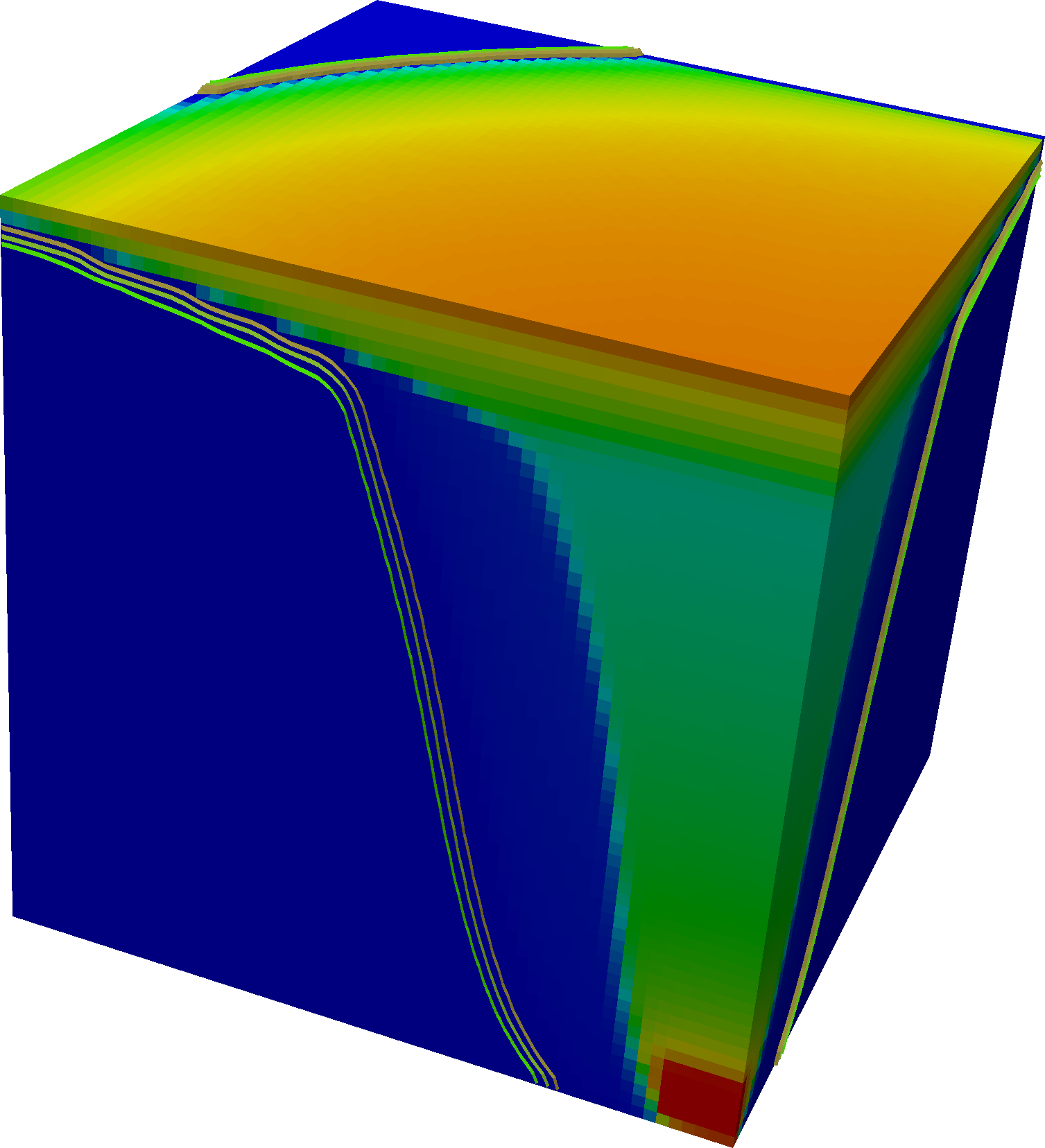}}
  \caption{Test case 3: \co phase saturation and molar fraction of
    dissolved \co in water (contour lines\\ for $x_\we^\nc=0.005$,
    $0.010$, $0.016$). Color scale ranges from $S_\nw = 0$ (blue) to
    $S_\nw = \max(S_\nw)$ (red).}
  \label{fig:11}
\end{figure*}

\section{Test case 3: \co injection in a fully water saturated domain (3D)}
\label{sec:8}
\begin{figure}
  \centering
  \includegraphics[width=.3\textwidth]{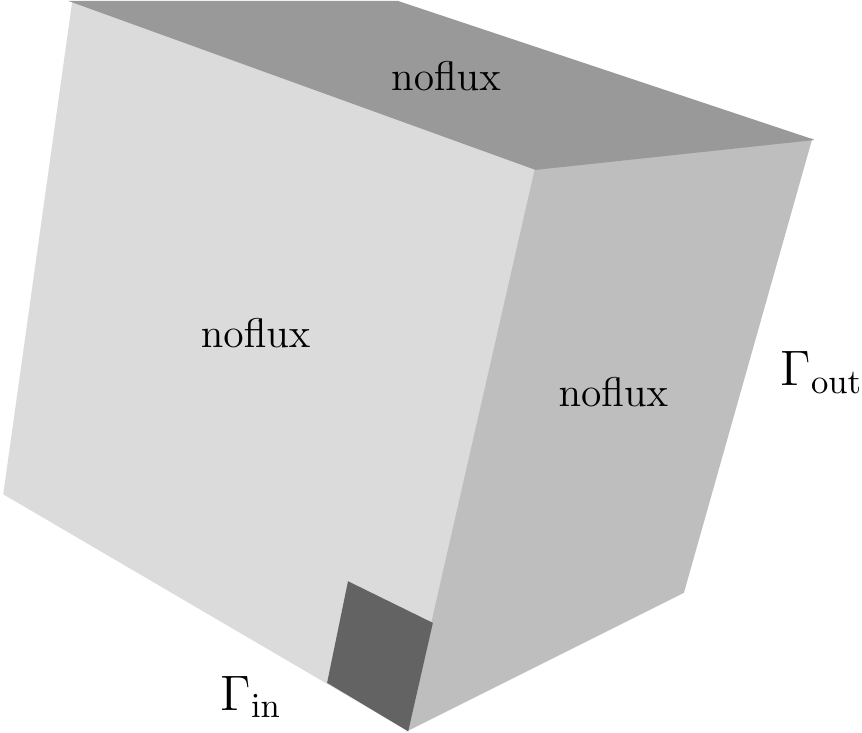}
  \caption{Test case 3: Domain setup}
  \label{fig:10}
\end{figure}
For test case 3, we use the same parameters and a very similar setup as
in test case 2 in Section \ref{sec:7}. The difference is that
we look at a 3D domain as shown in Figure \ref{fig:10}.

The domain is a cube with dimensions $\SI{100}{\metre} \times
\SI{100}{\metre} \times\SI{100}{\metre}$. For the computations a
structured grid with $60 \times 60 \times 60$ cells was used.

The results of test case 3 are shown in Figure \ref{fig:11}. As in test
case 2 each picture shows the \co phase saturation and the solubility
of \co in the water phase.

\section{Conclusion}
We suggest a new method to deal with the problem of disappearing
nonwetting phase in two-phase two-component flow simulations. We use
the nonwetting phase pressure and capillary pressure as primary
variables. This allows us to use the same variables for both the
monophasic and diphasic case, no switching of primary variables is
needed to treat the nonwetting phase appearance problem. For the
special case of CSS, we specify our choices for the necessary
constitutive relations.

We confirm our new choice of primary variables with numerical
simulations for different test cases in 2D and 3D. We simulate the
special case of \co injection in geological formations and took part
in the MoMas benchmark on multiphase flow, where hydrogen flow in
nuclear waste repositories was examined. All simulations are performed
in parallel and scale very well with the number of processes. In the
benchmark case our output corresponds very well to the results of
other groups.

Next we want to extend our simulations to a nonisothermal model and
use adaptive grid refinement. We want to use massive parallel computing
in order to simulate realistic CSS scenarios with very large domains
and long time spans.

\vspace{1cm}

{\small \noindent \textbf{Acknowledgments} We would like to thank Holger Class,
  Lena Walter and Melanie Darcis from the Department of Hydromechanics
  and Modeling of Hydrosystems at the University of Stuttgart for many
  fruitful discussions regarding the properties of \co-brine systems.
  This work was supported by the Baden-Württemberg Stiftung.}

{\small
\bibliographystyle{spmpsci}
\bibliography{main}
}
\end{document}